# Miniature GaAs disk resonators probed by a looped fiber taper for optomechanics applications

L. Ding[†a], P. Senellart[b], A. Lemaitre[b], S. Ducci[a], G. Leo[a], I. Favero[*a]

[a]Laboratoire Matériaux et Phénomènes Quantiques, Université Paris Diderot, CNRS, 10 rue Alice Domon et Léonie Duquet, 75013, Paris, France
[b]Laboratoire de Photonique et de Nanostructures, Route de Nozay, 91460 Marcousis, France

## ABSTRACT

GaAs disk resonators (typical disk size 5 μm × 200 nm in our work) are good candidates for boosting optomechanical coupling thanks to their ability to confine both optical and mechanical energy in a sub-micron interaction volume. We present results of optomechanical characterization of GaAs disks by near-field optical coupling from a tapered silica nano-waveguide. Whispering gallery modes with optical Q factor up to a few $10^5$ are observed. Critical coupling, optical resonance doublet splitting and mode identification are discussed. We eventually show an optomechanical phenomenon of optical force attraction of the silica taper to the disk. This phenomenon shows that mechanical and optical degrees of freedom naturally couple at the micro-nanoscale.

**Keywords:** GaAs, Disk Resonators, Optomechanics, Fiber Taper, Whispering Gallery Modes

## 1. INTRODUCTION

The coupling between light and mechanical oscillators has become a subject on its own with many complementary facets: applications of optomechanical systems are now evolving from sensing experiments to fundamental tests of quantum mechanics.[1,2] At the nano-scale, the optomechanical coupling increases thanks to a smaller optomechanical interaction volume and reduced mass of the mechanical oscillator.[3,4]

Tiny GaAs disk resonators, typically few microns in diameter and hundreds of nanometer in thickness, can be fabricated out of extra-pure material thanks to a combination of Molecular Beam Epitaxy (MBE) growth, advanced e-beam lithography and etching techniques. These systems are both mechanical oscillators and optical cavities. They combine many assets for nano-optomechanics experiments: high optical Q, nanoscale pg motional mass, sub-micron mode volume to confine both optical and mechanical energy. They are thus very good candidates for boosting phenomena relying on optomechanical coupling. These cavities are accessible through near-field evanescent spectroscopy, typically using an optical fiber taper.[5-8] In this article, we study such a GaAs nano-optomechanical disk resonator evanescently coupled to a tapered silica nano-waveguide. The optical properties of the disk cavity are characterized by taper transmission spectra. Disk whispering gallery modes are recorded as sharp transmission dips. Critical coupling, doublet resonance splitting in high Q gallery modes, and mode identification are discussed. Eventually we show that the fiber taper itself can serve as a sensitive sensor to detect minute optomechanical interaction in the disk-taper system. We use this principle to observe an optical gradient force.

[*]ivan.favero@univ-paris-diderot.fr
[†]lu.ding@univ-paris-diderot.fr

## 2. EXPERIMENT

GaAs disks used in our work are fabricated by e-beam lithography and subsequent wet etching method with typical disk size 5 µm × 200 nm (diameter × thickness).[9] The disk resonators are characterized by an evanescent fiber taper coupling technique shown in Fig. 1(b). In our experiment, a standard single mode silica optical fiber (10 µm core diameter and 125 µm cladding diameter) is adiabatically stretched down to ∼ 800 nm diameter using a microheater so that its evanescent field is made accessible to the surroundings. Furthermore, a micro-loop with a typical diameter of 70 µm is realized at the center part of the taper to improve spatial selectivity by creating a near-field optical "point-probe".[10] Using XYZ piezo-stages with 50 nm resolution, the looped fiber taper is positioned within the disk near-field to allow evanescent coupling (see Fig. 1(b) inset). Measurements of the taper transmission as a function of the taper-disk gap distance and input wavelength are performed using a fiber coupled external cavity tunable laser source (λ = 1500 – 1600 nm with linewidth ≤ 1 MHz).

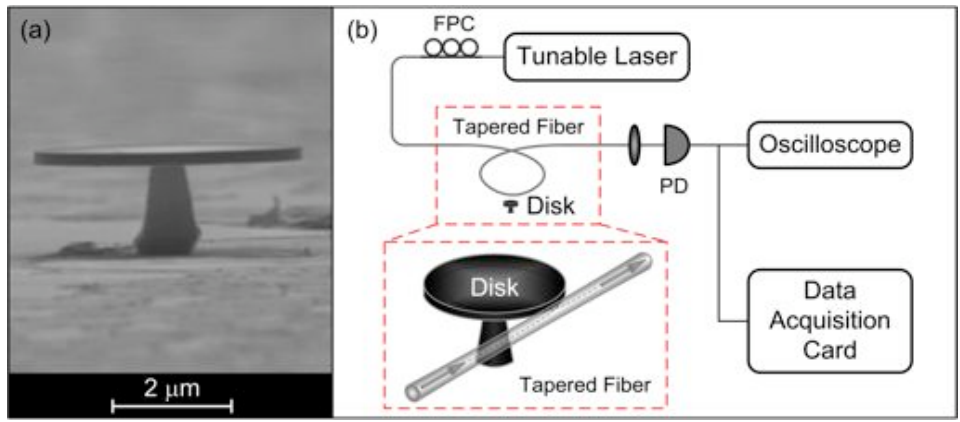

Figure 1. (a) Scanning Electron Microscope (SEM) picture of a GaAs disk resonator with thickness of 200 nm and diameter of 4.5 µm. (b) Schematic experiment of disk-taper transmission measurement utilizing the fiber taper coupling technique. FPC stands for fiber polarization controller, PD stands for photodetector. Inset is a close-up of the fiber-disk system.

In the looped taper, the polarization state is elliptical due to the birefringence effect induced by the curvature of the loop.[11] This ellipticity prevents perfect coupling to the TE- or TM-like whispering gallery modes (WGMs) of the disk. A fiber polarization controller (FPC) is used to adjust the polarization state in the loop, maximizing coupling to the selected WGM. The transmission signal is monitored on an oscilloscope and recorded on a computer through a data acquisition card.

## 3. SPECTROSCOPY OF DISK WHISPERING GALLERY MODES

### 3.1 Observation of critical coupling

When disk and taper are evanescently coupled, the disk WGMs appear as a series of transmission dips in the taper transmission spectrum. Figure 2(a) shows a single resonance of a disk WGM (disk diameter D = 11.8 μm) centered at 1546.8 nm measured when the taper is positioned laterally ~ 1.84 μm away from the edge of the disk. The gap distance is kept large in order to reduce taper loading effects.[6,8] Fitting the mode resonance to a Lorenzian gives a linewidth $\delta\lambda$ of 90 pm, indicating a Q factor of $1.7 \times 10^4$. Similarly in Fig. 2(b), we show the spectral response of the same resonance when the taper is positioned at the critical coupling distance (~ 280 nm from the edge of the disk). The presence of increased taper-disk coupling has modified the total loss rate of the resonant disk mode, yielding a loaded Q of $1.0 \times 10^4$. The resonance contrast, which refers to the depth of the on-resonance transmission dip, has also considerably increased from 8% to 57%.

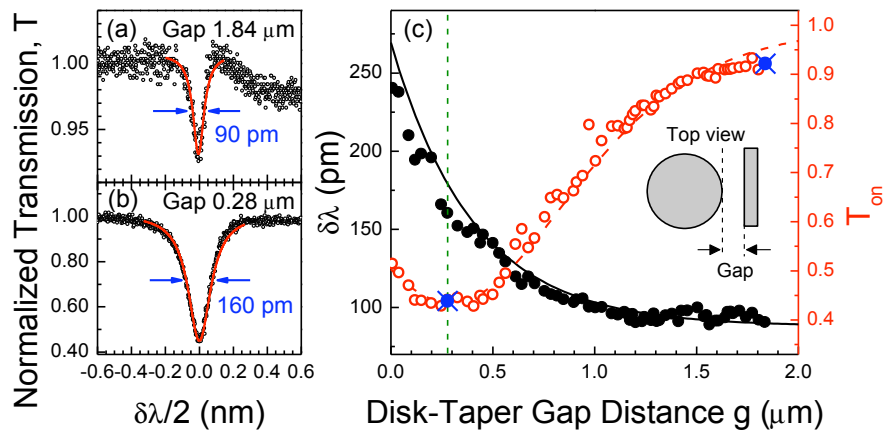

Figure 2. Normalized transmission spectra at (a) undercoupling and (b) critical coupling condition for an optical resonance centered at 1546.8 nm on a GaAs disk (D = 11.8 μm). (c) The transmission minimum $T_{on}$ (in red circle) and the linewidth $\delta\lambda$ (in black solid circle) as a function of gap distance together with fitting curves. Green dashed line separates the undercoupled (right) and overcoupled (left) regimes. Inset shows a schematic of the disk-taper configuration. Two blue points with a cross indicate gap distances at which positions the transmission spectra (a) and (b) are taken.

Figure 2(c) shows both the on resonance transmission minimum $T_{on}$ (in red circle) and the linewidth $\delta\lambda$ (in black solid circle) of the optical resonance as a function of the gap distance. The on-resonance transmission $T_{on}$ depending on the disk-taper optical coupling and loss rates can be expressed as,

$$T_{on} = T_{cc} + (1 - T_{cc}) \left( \frac{1 - \gamma_e/\gamma_i}{1 + \gamma_e/\gamma_i} \right)^2 \quad (1)$$

where $T_{cc}$ = 43% is the finite on resonance transmission at critical coupling induced by the polarization ellipticity in the loop. $\gamma_e$ is the extrinsic (taper-to-disk) coupling rate and $\gamma_i = (2\pi/\lambda_0)/Q_{in}$ is the intrinsic cavity loss rate.[12,13] $\lambda_0$ is the center wavelength of the resonance and $Q_{in}$ is the intrinsic Q factor. We expect $\gamma_e$ to vary exponentially with the gap distance, g, that is $\gamma_e(g) = \gamma_e(0)\exp(-\eta g)$, where $\gamma_e(0)$ is nominally the 'zero-gap' extrinsic coupling rate and $\eta$ is the decay constant. Fitting $T_{on}(g)$ (in red dashed curve), we extract the ratio of $\gamma_e(0)/\gamma_i$ = 2.062 and a decay constant equal to $\eta$ = 1/404 nm$^{-1}$ for the disk-taper system. In addition, the linewidth exponentially depends on the gap as,

$$\delta\lambda = \frac{\lambda_0}{Q_{in}} \left( 1 + \frac{\gamma_e}{\gamma_i} \right) \quad (2)$$

With the same value $\gamma_e(0)/\gamma_i$, $\eta$, and $Q_{in}$ = 1.7 x 10$^4$, the fit of $\delta\lambda(g)$ (in black curve) agrees well with experimental data.

### 3.2 Observation of high Q optical mode and doublet splitting

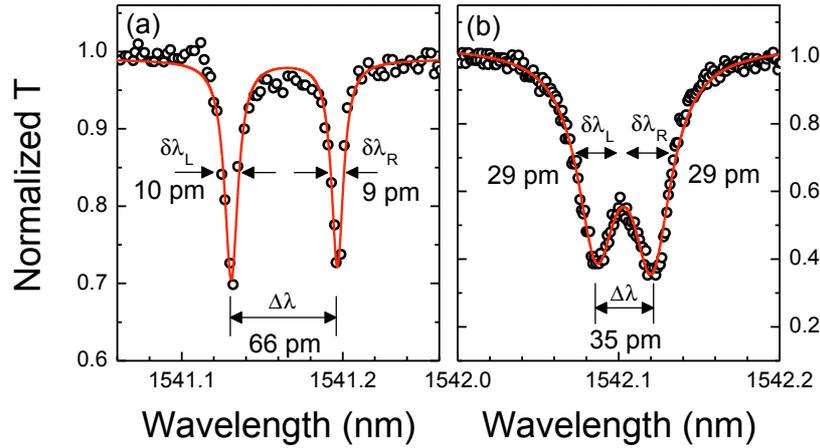

Figure 3. High resolution taper transmission spectra a showing WGM resonance doublet in the case of (a) $\Delta\lambda \gg \delta\lambda$ for the D = 9.8 μm disk and (b) $\Delta\lambda \approx \delta\lambda$ for the D = 7.0 mm disk. $\Delta\lambda$ and $\delta\lambda$ ($\delta\lambda_L$ and $\delta\lambda_R$) indicate the CW/CCW mode splitting and individual mode linewidth (left and right), respectively.

With high resolution wavelength scanning, we are able to observe high Q WGMs (up to 2 × 10$^5$) of our GaAs disk resonators. In many cases, the high Q optical mode resonance dip splits into a doublet[14], as shown in Fig. 3. The doublet splitting is attributed to the lifting of degeneracy of clockwise (CW) and counterclockwise (CCW) propagating WGMs in the disk resonator, as a consequence of Rayleigh scattering on disk surface roughness.[15-17] The rate at which photons are backscattered is quantified by the doublet splitting $\Delta\lambda$ while the rate at which photons are lost from the resonator is quantified by the intrinsic linewidth $\delta\lambda$ of the individual doublet modes. Fitting the transmission spectrum with a double Lorenztian, we find in Fig. 3(a) $\Delta\lambda$ = 66 pm and $\delta\lambda$ = 10 pm for a D = 9.8 µm disk. The individual mode loaded quality factor is $Q_{load} \equiv \lambda_0/\delta\lambda$ = 1.5 x $10^5$. In contrast to this example of $\Delta\lambda \gg \delta\lambda$, we obtain in Fig. 3(b) $\Delta\lambda$ = 35 pm and $\delta\lambda$ = 29 pm for the D = 7.0 µm disk, which gives an example of $\Delta\lambda \approx \delta\lambda$. The loaded Q factor of this resonance is $Q_{load}$ = 5.3 x $10^4$. We typically do not resolve the doublet splitting when Q is inferior to this value.

### 3.3 Optical WGMs identification

Figure 4(a) plots a broadband transmission spectrum of a GaAs disk resonator (R = 4.6 µm, t = 200 nm) with the looped taper in contact with the disk sidewall. The spectrum consists of a weak background oscillating with a period ~ 8 nm attributed to the quasi-ring structure of the looped taper[10] on top of which WGMs fine resonances are protruding (highest Q > 1 x $10^5$). WGMs are classified via the notation (p,m) where p and m are radial and azimuthal numbers respectively.[18] Although we are lacking the exact knowledge of the GaAs disk geometry, we can have an estimate of its mean radius with a precision of 300 nm by an optical microscope inspection. At this level of accuracy, the 200 nm thickness is guaranteed by the controlled MBE growth. Using finite element method (FEM) electromagnetic simulations[19], we compute the corresponding WGMs (p,m) wavelengths. This provides us with a spectral comb of modes, which is then shifted in block to overlap the measured optical resonances. The shift in wavelength to obtain overlap corresponds to a shift in disk radius inferior to 100 nm, consistent with the resolution of the optical microscope inspection. We then compute each mode radiative Q factor with the help of an home-made effective-index method code. The obtained Q factor helps confirming the mode identification procedure, by a rough comparison to the measured optical resonance width. Fig. 4(a) shows an example of such a TE (p,m) WGMs identification. Note that the mode comb sequence obtained from the simulations generally imposes a unique solution to the identification problem.

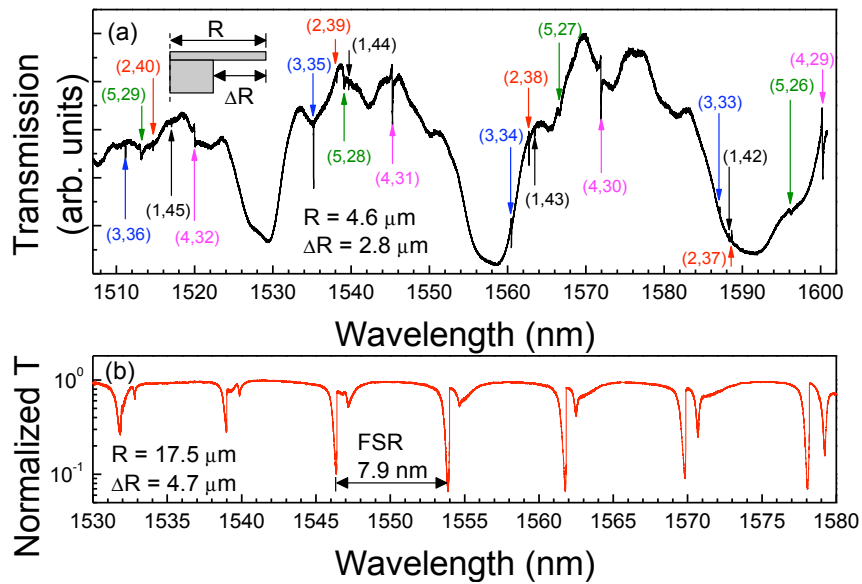

Figure 4. (a) Broad-band transmission spectrum, from 1500 to 1600 nm, of a GaAs disk resonator (R = 4.6 µm, t = 200 nm). WGMs are classified via the notation (p,m). The underetch distance $\Delta R$ = 2.8 µm, defined as radius difference between disk and pedestal, is superior to one half of the disk radius. (b) Broad-band transmission spectrum, from 1530 to 1580 nm, of a GaAs disk resonator (R = 17.5 µm, t = 200 nm) with underetch distance of 4.7 microns.

In an disk, the electric field of a p = 1 mode is strongly confined at the periphery of the disk while higher order radial modes distribute their energy density towards the center of the disk. When the pedestal is small enough (large underetch distance ΔR), high order p modes are supported by the disk. For example in Fig. 4(a) ΔR >1/2R, we are able to observed radial p modes with p equal 1 to 5.

In contrast Figure 4(b) shows a broadband transmission spectrum, from 1530 to 1580 nm, of a 17.5 μm radius GaAs disk resonator. The transmission was normalized to the spectral transmission of the fiber taper positioned far away from the disk. Here, the underetch ΔR = 4.7 μm is relatively small compared to the disk radius (ΔR << R).  Less  radial p WGMs are observed in the disk spectrum (only two type of radial p modes here). Indeed when the pedestal is large (ΔR << R), only low-order radial WGMs (p=1 or 2 typically) confined at the edge of the disk are preserved, while higher-order radial modes are quenched due to large optical losses induced on them by the presence of the pedestal. Taking the simple approximation $2\pi n_{eff} R = m\lambda_0$ to estimate the Free Spectral Range (FSR) with $\lambda_0$ = 1550 nm and $n_{eff}$ = 2.6 the effective index of the 200 nm GaAs slab TE mode, we find FSR = $\lambda_0^2/(2\pi n_{eff} R)$ = 8.4 nm which agrees well with the experimental value of 7.9 nm observed in Fig 4(b).

## 4. OPTICAL FORCE ATTRACTION OF THE TAPERED FIBER

Theoretical prediction of an optical force acting on evanescently coupled optical waveguides[20] or resonant structures[21] was followed by its recent observation in a disk-taper system in Ref. 12. In this work the authors showed that milliwatt optical power can produce micrometer-scale displacement of a movable input waveguide (a silica optical-fiber taper) evanescently coupled to a high-Q $SiN_x$, disk resonator (D = 44.8 μm and t = 253 nm). The displacement was caused by a cavity-enhanced optical dipole force pulling on the waveguide.

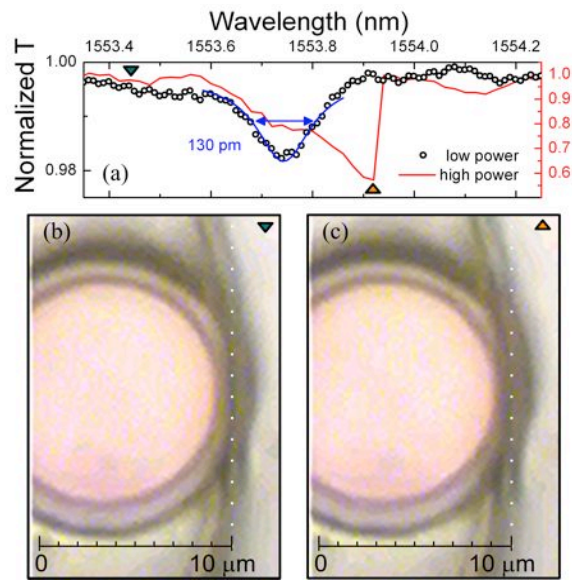

Figure 5. (a) Low optical input power (black circle) transmission spectrum of a WGM (Q ~ 1.2 x 10$^4$) of the disk (R = 5.7 μm, t = 200 nm). High optical input power (red curve) transmission spectrum of the same WGM showing thermo-optic distortion and a change in the on resonance transmission minimum. The green (yellow) marker indicates the transmission level in (b) and (c). (b) An optical micrograph of the taper-disk coupling region taken off resonance at high power. White dotted line indicates the original position of the taper inner edge. (c) An optical micrograph taken on resonance at high power, showing a significant displacement of the taper waveguide towards the disk. White dotted line is here for reference at the same position as in (b).

The GaAs disk resonator shown in Fig. 5 has radius R = 5.7 μm and thickness t = 200 nm. Optical power is injected into the disk through evanescent coupling from the looped taper. An optical micrograph of the fiber-taper waveguide (diameter ~ 1 μm) in the near-field of the GaAs disk is shown in Fig. 5(b). Figure 5(a) shows the transmission spectra of a WGM mode (Q ~ 1.2 x 10$^4$) at low and high power, * and * μW falling on the photodetector respectively. At low power the symmetric transmission resonance indicates that the disk-taper system behaves linearly. At higher power, thermo-optic effects within the GaAs result in an asymmetric resonance in the scanned transmission spectrum.[12,22,23] In addition, the on-resonance transmission significantly decreases with increasing input power. When scanning-up the wavelength towards the resonance minimum, the taper moves towards the disk. This effect is shown in Figs. 5(b) and 5(c). The displacement is as large as about 500 nm, easily visible with the optical microscope integrated in our experiment. It is the result of the cavity-enhanced optical dipole force. At resonance, the electric field amplitude in the disk is amplified. The taper is positioned in the evanescent tail and becomes polarized. The dipole force attracts the taper towards the disk. We have estimated that an optical force inferior to 1 pN is responsible for this displacement. The looped taper is mechanically very soft in our experiment, making it sensitive to extremely small forces. Note that if a single polarizable atom is easily trapped and displaced by a moderate power laser beam, the amplification of the optical force by the disk resonance is here needed to produce visible displacement of the macroscopic fiber. The here observed phenomenon shows that mechanical and optical degrees of freedom tend to couple naturally at the micro-nanoscale. It also proves the relevance of optical methods for actuation of micro or nanomechanical switches.

## 5. CONCLUSIONS

We present a semiconductor nano-optomechanical system consisting of a GaAs disk resonator coupled to a tapered silica nano-waveguide. The optical properties of the disk cavity are characterized by taper transmission spectrum. Whispering gallery modes with optical Q factor up to a few $10^5$ are observed. Critical coupling, resonance doublet splitting and modes identification are discussed. As an example of optomechanical interaction, we show a phenomenon of optical force attraction of the silica waveguide towards the disk, with a resulting micron-scale displacement.

## ACKNOWLEDGEMENT

This work was supported by C-Nano Ile de France NAOMI project.

system for application to cavity quantum electrodynamics," Phys. Rev. Lett. 91(4), 043902 (2003).